# Knotted spacetime electromagnetic vortex unlinking and unknotting with vector and scalar reconnections and field twist compensation

Jordan M. Adams

Riverside Research, Beavercreek, Ohio 45431, United States.

**Abstract:** Optical vortex knots have been realized in monochromatic laser beams, but monochromatic fields are stationary and their topology is frozen. Here we show that knotted spatiotemporal vortices, whose phase singularities form closed loops in spacetime, undergo topology-changing reconnections with free-space propagation. When null lines of different vector components unlink, the electric spin, magnetic spin, linear momentum, and electromagnetic helicity densities, each built from a specific pair of field components, twist to exactly compensate the change in linking number. This compensation is enforced by the argument principle where the total for each component pair, combining mutual phase twist, geometric linking, and open-line threading, vanishes identically and remains exactly zero through all reconnection events.

In classical fluids, the helicity $\mathcal{H} = \int \mathbf{v} \cdot \boldsymbol{\omega}\, d^3x$ is approximately conserved through reconnections [1,2], with the Călugăreanu–White–Fuller theorem decomposing helicity into the topological and geometric properties of linking, writhe, and twist [3,4]. Magnetic helicity obeys an analogous approximate conservation in plasmas [5,6]. In both systems, however, conservation degrades with viscosity or resistivity and is exact only in the dissipationless limit.

Optical vortex lines, the phase singularities of complex wave fields [7], can be knotted and linked in monochromatic beams [8-10], and carry topological textures such as polarization skyrmions [11-12]. Monochromatic fields, however, are stationary, and their vortex topology cannot change. Berry & Dennis proved that the phase twist of a closed dislocation loop equals the dislocation strength threading through the loop [7], applying Cauchy's argument principle to wave singularities, and Dennis connected this to the Călugăreanu relation, Lk = Tw + Wr, for phase ribbons, which shows that the writhe must compensate changes in twist and vice-versa [13]. These results establish identities for a single scalar field at a fixed instant but do not address conservation through topology-changing reconnection.

Spatiotemporal optical vortices, polychromatic beams whose phase singularities exist as lines in $(x, y, t)$ at each propagation plane $z$, have finite extent and can undergo reconnection as $z$ changes [14, 15]. Spatiotemporal vortex knots were first constructed numerically from Milnor polynomials and shown to unknot through reconnection upon propagation [16], and have recently been demonstrated experimentally [17]. These reconnections change the knot topology, but Maxwell's equations couple all six electromagnetic components $(E_x, E_y, E_z, B_x, B_y, B_z)$, each with its own family of null lines. Whether exact conservation laws govern this process, analogous to helicity conservation in fluids but without the approximation, has remained an open question.

Here we show that for pairs of field components, the combination of mutual phase twist, geometric linking, and open-line threading, vanishes identically and remains exactly zero through all reconnection events. The mutual phase twist between component pairs is encoded in physical observables including electric spin, magnetic spin, linear momentum, and electromagnetic helicity. When null lines of different components unlink, the Stoke vectors defining the observables acquire twist to exactly compensate the change in linking number, providing a measurable signature of exact topological conservation in electromagnetic reconnection.

Spatiotemporal fields like these knotted spatiotemporal vortices hold promise in several applications such as free-space communications, but the complicated evolution with propagation make this an undeveloped area. We believe this work provides an initial theoretical foundation for knotted spatiotemporal vortices and their use in future applications. Additionally, fluid reconnections processes are clouded by dissipation and nonlinearities. These results from deterministic electromagnetic propagation may give insight into fluid reconnections.

## Results

In two dimensions, Cauchy's argument principle states that the phase winding of a complex function along a closed contour counts the enclosed zeros [18]. Applied to electromagnetic vortices, this means the phase twist of a vortex loop counts the singularities that thread through the loop [7]. When the contour is a closed curve in three-dimensional space, winding counts become linking numbers [19]. If $\Phi_b(\mathbf{r})$ is a complex field whose zero set consists of $N$ closed curves $\gamma_b^n$, $n: \{1,2,\ldots N\}$, then the total twist or phase winding of $\Phi_b$ along any closed curve $\gamma_a$ of another field $\Phi_a(\mathbf{r})$ is

$$\text{tw}(b \mid a) \equiv \sum_n \frac{1}{2\pi} \oint_{\gamma_a^n} d\,[\arg \Phi_b], \tag{1}$$

and the total Gauss linking number [20],

$$\text{Lk}(\gamma_a, \gamma_b) = \sum_n \sum_m \frac{1}{4\pi} \oint_{\gamma_a^m} \oint_{\gamma_b^n} \frac{(\mathbf{r}_1 - \mathbf{r}_2) \cdot (d\mathbf{r}_1 \times d\mathbf{r}_2)}{|\mathbf{r}_1 - \mathbf{r}_2|^3}. \tag{2}$$

Applied to the six electromagnetic components $\Phi_a \in \{E_x, E_y, E_z, B_x, B_y, B_z\}$, each with null curves $\gamma_a^n$, this yields 36 mutual twists $\text{tw}(b \mid a)$ that we collect into a $6 \times 6$ winding matrix $\text{Tw}_{ab} = \text{tw}(b \mid a)$, where row $a$ selects the null curves of field $\Phi_a$ and column $b$ the sum of accumulated twist value of field $\Phi_b$ along $\Phi_a's$ null lines. For example, $tw(E_x, E_y)$ measures how many multiplies of $2\pi$ the phase of $E_x$ winds around all of $E_y$'s closed null lines. Together with the Gauss linking matrix $\text{Lk}_{ab} = \text{Lk}(\gamma_a, \gamma_b)$, it defines a conservation matrix based on the argument principle

$$\mathcal{C} = \mathbf{Tw} + \mathbf{Lk} + \mathcal{T}, \tag{3}$$

where $\mathcal{T}$ is the threading matrix. Not all null lines close into loops. Components whose null set includes open lines, singularities that extend through the domain without closing, contribute a threading number $\mathcal{T}_{ab}$ equal to the number of times these open lines pierce any surface bounded by the closed loops [20]. We demonstrate numerically below that $\mathcal{C}_{ab} = 0$ for all $a, b$. The closed-loop contribution $C^l_{ab} = \text{Tw}_{ab} + \text{Lk}_{ab}$ is independently conserved under propagation, but non-zero, with $\mathcal{T}_{ab}$ providing the constant offset.

Each entry of **Tw** encodes a distinct physical observable, because familiar electromagnetic quantities are built from relative phases between field components, and the argument principle decomposes each relative phase into winding-matrix entries. Since $\arg(\Phi_a^* \Phi_b) = \arg\Phi_b - \arg\Phi_a$, the winding along any closed curve $\gamma$ separates as

$$\frac{1}{2\pi} \oint_\gamma d\,[\arg(\Phi_a^* \Phi_b)] = \text{tw}(b \mid \gamma) - \text{tw}(a \mid \gamma). \tag{4}$$

The twist of a bilinear observable $\arg(\Phi_a^* \Phi_b)$ at a null line is the number of $2\pi$ phase windings accumulated along the null line. Evaluating on a single null of $\Phi_a$ gives

$$\frac{1}{2\pi} \oint_{\gamma_a^n} d\,[\arg(\Phi_a^* \Phi_b)] = \text{tw}(b \mid a)|_{\gamma_a^n} - \text{tw}(a \mid a)|_{\gamma_a^n}, \tag{5}$$

where $\text{tw}(a \mid a)|_{\gamma_a^n}$ is the self-twist on null line $\gamma_a^n$. Each pair of components $(a, b)$ defines a Poincaré sphere through the Stokes parameters $S_1 = |\Phi_a|^2 - |\Phi_b|^2$, $S_2 = 2\,\text{Re}(\Phi_a^* \Phi_b)$, $S_3 = 2\,\text{Im}(\Phi_a^* \Phi_b)$, with $\arg(\Phi_a^* \Phi_b) = \arctan(S_3/S_2)$ as the azimuthal coordinate. Eq (5) counts how many times this Stokes vector wraps the sphere along a null line. For $E$–$E$ pairs the sphere describes the polarization state, and the electric spin density is its $S_3$ component, $2\,\text{Im}(E_i^* E_j)$ (Table 1). For $E$–$B$ pairs, $S_3 = 2\,\text{Im}(E_i^* B_i)$ gives the helicity density and $S_2 = 2\,\text{Re}(E_j^* B_k)$ gives the momentum density or Poynting vector. For $B$–$B$ pairs, $S_3$ gives the magnetic spin. Each observable's twist is constrained by the equation 5 argument principle matrix. For example, the electric spin phase angle relation is $\frac{1}{2\pi} \oint_{\gamma_i^n} d\,[\arg(E_i^* E_j)] = Lk_{E_i,E_i} - Lk_{E_i,E_j} + \mathcal{T}_{E_i,E_i} - \mathcal{T}_{E_i,E_j}$.

*Table 1: Decomposition of electromagnetic observables into winding-matrix entries via equation (5). Each observable has phase angle $\arg(\Phi_a^* \Phi_b)$, whose winding on $\gamma_a$ is a difference of mutual and self-twists.*

| Observable | Phase angle | Phase winding on $\gamma_i$ | Twist relation |
|---|---|---|---|
| Electric spin | $\arg(E_i^* E_j)$ | $\text{tw}(E_j \mid E_i) - \text{tw}(E_i \mid E_i)$ | $Lk_{E_i,E_i} - Lk_{E_i,E_j} + \mathcal{T}_{E_i,E_i} - \mathcal{T}_{E_i,E_j}$ |
| Momentum | $\arg(E_i^* B_k)$ | $\text{tw}(B_k \mid E_j) - \text{tw}(E_j \mid E_j)$ | $Lk_{E_i,E_i} - Lk_{E_i,B_j} + \mathcal{T}_{E_i,E_i} - \mathcal{T}_{E_i,B_j}$ |
| EM helicity | $\arg(E_i^* B_i)$ | $\text{tw}(B_i \mid E_i) - \text{tw}(E_i \mid E_i)$ | $Lk_{E_i,E_i} - Lk_{E_i,B_i} + \mathcal{T}_{E_i,E_i} - \mathcal{T}_{E_i,B_i}$ |
| Magnetic spin | $\arg(B_i^* B_j)$ | $\text{tw}(B_j \mid B_i) - \text{tw}(B_i \mid B_i)$ | $Lk_{B_i,B_i} - Lk_{B_i,B_j} + \mathcal{T}_{B_i,B_i} - \mathcal{T}_{B_i,B_j}$ |

To validate the theory, we construct spatiotemporal knotted vortices based on Milnor polynomials [10, 21]. We start with a scalar seed field

$$\Psi(\mathbf{r}, t) = [\alpha\,\tilde{p}^m + \tilde{q}^n]\,e^{-\frac{r^2}{w_0^2}}\,e^{-\frac{t^2}{T_0^2}}\,e^{-i\omega_0 t}, \tag{6}$$

with dimensionless variables

$$\tilde{p} = \frac{r^2 - r_0^2}{w_0^2} + i\frac{t}{T_0}, \quad \tilde{q} = \frac{i(x + iy)}{w_0}, \tag{7}$$

and use common lab parameters for ultrashort THz pulses: $r_0 = 1.0$ mm, $w_0 = 2.0$ mm, $T_0 = 2.0$ ps, $\alpha = 2^{2m-n}$, $\omega_0 = k_0 c$ with $k_0 = 10$ mm$^{-1}$. The zero set of the polynomial factor is a $(2m, n)$ torus knot or link in $(x, y, t)$. Varying $(m, n)$ produces the unknot, Hopf link, trefoil, and higher torus knots.

The seed $\Psi$ is embedded in a vector potential with near-circular polarization and a small longitudinal tilt: $\hat{\mathbf{A}} = (1/\sqrt{2},\ i/\sqrt{2},\ \epsilon_{az})\,\hat{\Psi}$, where $\epsilon_{az} = 0.2$. The tilt breaks the degeneracy between transverse and longitudinal components, giving each an independent null-line topology. The six field components are obtained from $\hat{E}_i = -i\omega\,(\hat{A}_i - k_i\,\mathbf{k}\cdot\hat{\mathbf{A}}/k^2)$, $\hat{\mathbf{B}} = (\mathbf{k}/\omega) \times \hat{\mathbf{E}}$, and propagated using the angular spectrum method with $k_z = \sqrt{(\omega/c)^2 - k_x^2 - k_y^2}$.

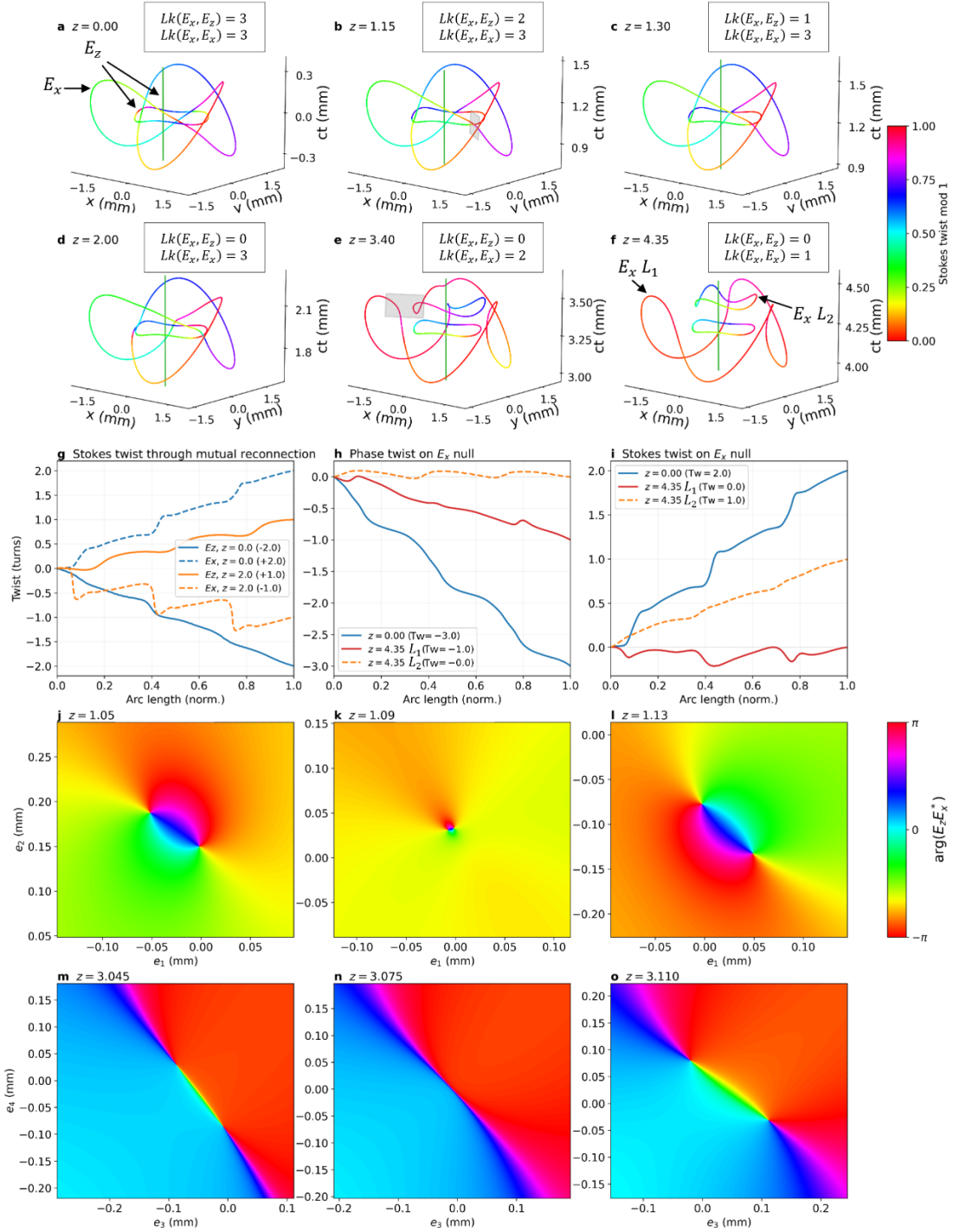

*Figure 1*: Null-line topology through mutual and self-reconnection. **a–c,** $E_x$ and $E_z$ null lines unlinking with propagation; the open $E_z$ null line threads through all closed loops. Curves coloured by cumulative Stokes twist. **d–f,** Further propagation: the $E_x$ trefoil unknots via self-reconnection. **g,** Electric spin Stokes twist $\arg(E_z^* E_x)$ along $E_x$ and $E_z$ null lines before and after mutual reconnection. **h,** Self phase twist along the $E_x$ null line before and after self-reconnection. **i,** Electric spin Stokes twist along $E_x$ null line, same z-values as **h**. **j–l,** Phase of $\arg(E_z^* E_x)$ on cross-sections through the $E_x$–$E_z$ crossing (grey square region in **b**). **m–o,** Phase of $\arg(E_z^* E_x)$ cross-sections through the $E_x$ self-reconnection (grey square region in **e**).

The trefoil beam $f = p^2 + q^3$ is tracked through free-space propagation (Fig. 1). At small $z$, the $E_x$ null line is a trefoil knot linked three times with the $E_z$ inner loop, while the open $E_z$ null line threads through all closed curves (Fig. 1a). With propagation, mutual reconnection reduces the cross-vector linking from 3 to 0 (Fig. 1a–c). The Stokes twist $\arg(E_z E_x^*)$ accumulated along each null line redistributes to compensate (Fig. 1g). At each reconnection the twist jumps by exactly one unit, as the argument principle enforces integer steps. Cross-sections perpendicular to the null-line crossing (Fig. 1j–l) show the electric spin Stokes twist $\arg(E_z E_x^*)$. At the $E_x$ null line, $E_x$ vanishes and this phase is undefined. Circling the null lines, the $E_x$–$E_z$ Stokes vector wraps the Poincaré sphere once, giving skyrmion number [12] $N_{\text{sky}} = (4\pi)^{-1} \int \mathbf{S} \cdot (\partial_u \mathbf{S} \times \partial_v \mathbf{S})\, du\, dv = 1$. At the reconnection the two skyrmion lines intersect, and a skyrmions dipole annihilates ($N_{\text{sky}} \to 0$) and reforms afterward. At $z \approx 3$, the $E_x$ trefoil itself reconnects, splitting into two daughter loops (Fig. 1d–f) with in its own linking and corresponding redistribution of phase twist and Stokes twist (Fig. 1h,i), as well as skyrmion dipole annihilation (Fig. 1m-o).

Tracking all six components and computing the mutual linking, mutual twist, self-phase twist, self-linking ($Lk_s$), and open-line threading (see Methods) at every propagation distance confirms the full equation 3 conservation matrix (Fig. 2). Additionally, the mutual linking between null lines within one vector component is plotted ($Lk_m$) and receives a double count from the double sum in equation 2. The $E_x$ self-conservation $\mathcal{C}(E_x, E_x) = 0$ holds exactly through both reconnection events (Fig. 2a). For $E_z$, which has two loops (larger outer loop not shown in Fig. 1) the closed-loop part $\mathcal{C}^l = -2$ is offset by the open-line threading $\mathcal{T}(E_z, E_z) = +2$, giving $\mathcal{C}(E_z, E_z) = 0$ (Fig. 2b). The mutual conservation $\mathcal{C}(E_x, E_z) = 0$ is also verified at every $z$, as Lk drops Tw compensates exactly (Fig. 2c). Summing all entries of $\mathcal{C}$ gives the total

$$\mathcal{H} = \sum_{a,b} \mathcal{C}_{ab} = \mathcal{H}_l + \mathcal{H}_{\text{th}} = 0, \tag{8}$$

with $\mathcal{H}_l = -24$ and $\mathcal{H}_{\text{th}} = +24$ for the trefoil (Fig. 2d).

While we only visualized the $E_x, E_z$ unlinking, each vector component pair undergoes separate transitions. Table 2 gives a better perspective for each vector component pair, showing results at $z = 0$ and $z = 5$. Since the trefoil topology is similar for $E_x, E_y, B_x, B_y$ and the loop and thread topology similar for $E_z, B_z$, there are several degenerate values for twist, linkage, and thread count. The closed-loop part $\mathcal{C}^l$ is non-zero only when a longitudinal component is involved ($\mathcal{C}^l = -2$ for pairs involving longitudinal polarizations), precisely cancelled by the threading $\mathcal{T}$ of the open $E_z$ and $B_z$ null lines. Each mutual entry corresponds to a different observable's Stokes twist (Table 1). As the $E_x$–$E_z$ linking drops from 3 to 0, the electric spin Stokes twist compensates exactly. The same holds independently for the linear momentum ($E_x, B_z$), electromagnetic helicity ($E_z, B_z$) and ($E_x, B_x$), and magnetic spin ($B_x, B_z$). The longitudinal helicity ($E_z, B_z$) has non zero threading due to the longitudinal components open lines, while the transverse helicity has zero threading from the purely closed loop trefoil curves.

**Table 2:** *Conservation matrix $\mathcal{C}$ for the trefoil beam at $z = 0$ and $z = 5$, showing redistribution of linkage and twist while each entry remains independently conserved.*

| | | | | $z = 0$ | | | $z = 5$ | | |
|---|---|---|---|---|---|---|---|---|---|
| Category | Component 1 | Component 2 | $\mathcal{C}$ | $tw(b\|a)$ | Lk | $\mathcal{T}$ | $tw(b\|a)$ | Lk | $\mathcal{T}$ |
| Transverse, Longitudinal | $E_x, E_y, B_x, B_y$ | $E_z, B_z$ | 0 | $-5$ | $+3$ | $+2$ | $-2$ | 0 | $+2$ |
| Longitudinal, Transverse | $E_z, B_z$ | $E_x, E_y, B_x, B_y$ | 0 | $-3$ | $+3$ | 0 | 0 | 0 | 0 |
| Category | Pair/Component | | $\mathcal{C}$ | $tw(b\|a)$ | Lk | $\mathcal{T}$ | $tw(b\|a)$ | Lk | $\mathcal{T}$ |
| Transverse self | $E_x, E_x, E_y, E_y, B_x, B_x, B_y, B_y$ | | 0 | $-3$ | $+3$ | 0 | $-1$ | $+1$ | 0 |
| Longitudinal self | $E_z, E_z, B_z, B_z$ | | 0 | $-3$ | $+1$ | $+2$ | $-3$ | $+1$ | $+2$ |
| Longitudinal helicity | $(E_z, B_z)$ | | 0 | 0 | $-2$ | $+2$ | 0 | $-2$ | $+2$ |
| Transverse spin/helicity | $(E_x, E_y), (E_x, B_x), (E_y, B_y)$ | | 0 | $-4$ | $+4$ | 0 | $-2$ | $+2$ | 0 |
| Transverse magnetic spin | $(B_x, B_y)$ | | 0 | $-5$ | $+5$ | 0 | $-2$ | $+2$ | 0 |
| Longitudinal Poynting | $(E_x, B_y), (E_y, B_x)$ | | 0 | $-7$ | $+7$ | 0 | $-7$ | $+7$ | 0 |

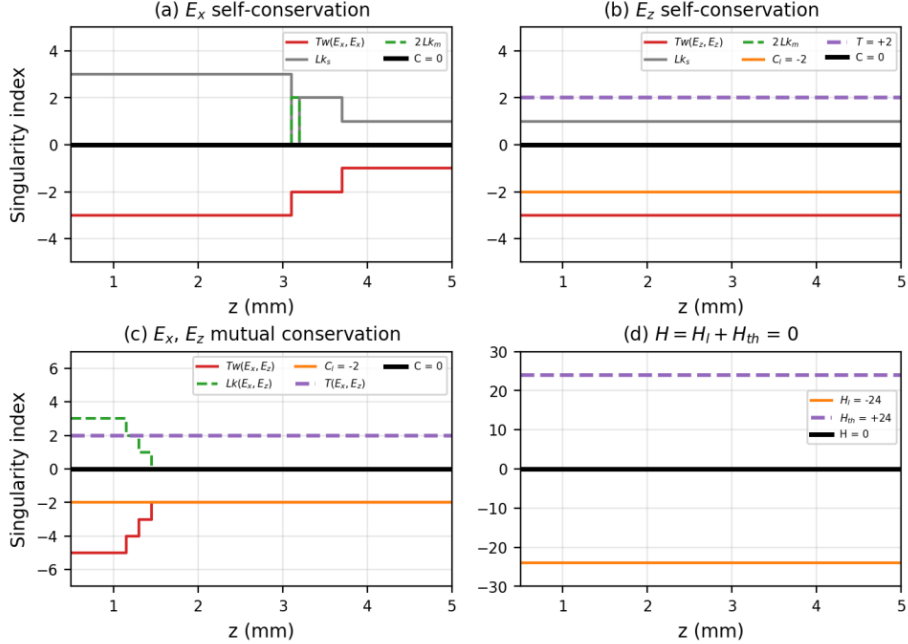

**Figure 2**: $\mathcal{C} = 0$ *verified through propagation.* **a,** $E_x$ self-conservation: $\mathcal{C} = Tw(E_x, E_x) + Lk_s + 2\,Lk_m = 0$ (black). **b,** $E_z$ self-conservation: the closed-loop part $\mathcal{C}_l = -2$ (orange) is offset by threading $\mathcal{T} = +2$ (purple dashed), giving $\mathcal{C} = 0$ (black). **c,** $E_x$–$E_z$ mutual conservation: $\mathcal{C}_l = tw(E_z|E_x) + Lk = -2$ (orange); $\mathcal{T} = +2$ (purple dashed); $\mathcal{C} = 0$ (black). **d,** Total: $\mathcal{H}_l = -24$ (red), $\mathcal{H}_{th} = +24$ (orange), $\mathcal{H} = 0$ (black).

We also investigated additional knot types by varying the polynomial degrees $(m, n)$ (Table 3). The results show that the closed-loop contribution is $\mathcal{H}_l = -12m$, which depends only on $m$ and is independent of the knot type $n$. The unknot, Hopf link, trefoil, and unlink, which are topologically distinct, all give $\mathcal{H}_l = -24$ when $m = 2$. In every case the threading $\mathcal{H}_{th} = +12m$ cancels exactly, giving $\mathcal{H} = 0$.

**Table 3:** *Threading for Milnor polynomials $p^m + q^n$. $\mathcal{H}_l$ is the closed-loop contribution (linking + twist), $\mathcal{H}_{th}$ is the open-line threading, and $\mathcal{H} = \mathcal{H}_l + \mathcal{H}_{th} = 0$ in every case.*

| Polynomial | $(m,n)$ | Topology | $\mathcal{H}_l$ | $\mathcal{H}_{th}$ | $\mathcal{H}$ |
|---|---|---|---|---|---|
| $p + q$ | (1,1) | unknot | $-12$ | $+12$ | 0 |
| $p^2 + q$ | (2,1) | unlink | $-24$ | $+24$ | 0 |
| $p^2 + q^2$ | (2,2) | Hopf link | $-24$ | $+24$ | 0 |
| $p^2 + q^3$ | (2,3) | trefoil knot | $-24$ | $+24$ | 0 |
| $p^3 + q^3$ | (3,3) | 3-component link | $-36$ | $+36$ | 0 |

The scaling $\mathcal{H}_l = -12m$ follows because $m$ controls the number of open null lines created by the factor $p^m$, so the total threading scales with $m$ but is independent of the braiding pattern set by $q^n$. The threading is also insensitive to chirality: replacing $q = i(x + iy)$ with its conjugate $\bar{q} = i(x - iy)$ produces the mirror-image trefoil, but the threading numbers and $\mathcal{H}$ are unchanged.

## Discussion

For a single scalar field, Berry & Dennis showed that the phase twist of a dislocation loop equals the threading strength [7], and Dennis connected this to the Călugăreanu decomposition [13,22]. Our conservation matrix extends this from a single framed curve to six families of curves framed by one another's phases, and from a static identity to a dynamical conservation law through reconnection, the key distinction being that reconnection can occur only in spatiotemporal pulses, not in monochromatic fields.

Fluid helicity degrades through viscous dissipation and magnetic helicity through resistive diffusion; in contrast, the invariants here hold at every propagation distance because they rest on the integer-valued argument principle rather than on an energy inequality. The matrix formulation is not specific to electromagnetism. It applies to any system with multiple complex scalar fields whose null lines form closed curves.

Spatiotemporal vortex reconnections [14], and spatiotemporal vortex knots have been demonstrated experimentally [17]. The conservation law predicts that any such closed loop reconnection event must be accompanied by an exactly compensating twist in the electromagnetic observables, a testable signature accessible through measurement of the field components.

## Conclusion

This work showed that spatiotemporal vortex knots undergo topology-changing reconnections with free-space propagation. When null lines of different vector components unlink, the electric spin, magnetic spin, linear momentum, and electromagnetic helicity densities, each built from a specific pair of field components, twist to exactly compensate the change in linking number. This compensation is enforced by the argument principle: the total for each component pair, combining mutual phase twist, geometric linking, and open-line threading, vanishes identically and remains exactly zero through all reconnection events. We believe these results will be beneficial for future applications that use spatiotemporal vortex knots and may benefit understanding of fluid reconnection processes.

## Methods

*Field construction and propagation.*

The Milnor polynomial (7) is multiplied by a Gaussian envelope $g = \exp[-(r/w_0)^2 - (t/T_0)^2]$ and Fourier-transformed to the spectral domain $(\mathbf{k}, \omega)$. Propagation to distance $z$ applies the exact transfer function $H = \exp(ik_z z)$ with $k_z = \sqrt{(\omega/c)^2 - k_x^2 - k_y^2}$; evanescent components ($k_z^2 < 0$) are set to zero. The six field components are obtained from $\hat{E}_i = -i\omega\,(\hat{A}_i - k_i\,\mathbf{k}\cdot\hat{\mathbf{A}}/k^2)$, $\hat{\mathbf{B}} = (\mathbf{k}/\omega) \times \hat{\mathbf{E}}$. The computation uses $N_x = N_y = N_t = 384$ grid points in space and time.

*Null-line detection and tracing.*

Null lines of each component are detected on the 3D $(x, y, t)$ grid by finding intersections of $\text{Re}\,\Phi = 0$ and $\text{Im}\,\Phi = 0$ contours on 2D slices at multiple time values. Seed points are refined by Newton correction and connected into continuous curves by predictor–corrector marching with step size $ds = 0.008$. For polynomials with $m \geq 2$, closely spaced null sheets cause standard Newton correction to hop between sheets; we use a constrained corrector that projects onto the plane perpendicular to the local tangent direction: $[\nabla\text{Re}\,\Phi;\ \nabla\text{Im}\,\Phi;\ \hat{\mathbf{t}}^T]\,\delta\mathbf{r} = [-\text{Re}\,\Phi;\ -\text{Im}\,\Phi;\ 0]$. Curves are validated by confirming closure (endpoint gap $< 0.05$) and checking that the field residual along each curve remains below $10^{-5}$.

*Linking numbers.*

The mutual linking number $\text{Lk}(a, b)$ is computed from the Gauss double integral $\text{Lk} = (4\pi)^{-1} \oint \oint (\mathbf{r}_1 - \mathbf{r}_2) \cdot (d\mathbf{r}_1 \times d\mathbf{r}_2)/|\mathbf{r}_1 - \mathbf{r}_2|^3$. The self-linking number $Lk_s$ is the Gauss linking integral between the curve and a parallel-transport push-off with holonomy correction, at distance $\epsilon = 0.05$ mm.

*Twist.*

The mutual twist $\text{tw}(b \mid a)$ is computed by evaluating the complex field $\Phi_b$ at each point of curve $\gamma_a$ via trilinear interpolation on the 3D grid and summing the wrapped phase increments $\Delta\phi_i = \arg(\Phi_{b,i+1}/\Phi_{b,i})$. The total twist is $\text{tw} = (2\pi)^{-1} \sum_i \Delta\phi_i$. The self phase twist is computed identically, but evaluating $\Phi_a$ along the push-off curve (at the same $\epsilon = 0.05$ mm used for $Lk_s$) rather than along $\gamma_a$ itself, where the field vanishes.

*Open-line threading.*

Open null lines, which do not close into loops, are detected by scanning for phase singularities (winding $> \pi$) near the pulse centre in each $(x, y)$ time slice. The threading number $\mathcal{T}_{ab}$ is computed by fan-triangulating a surface from the centroid of each closed curve and counting signed intersections of the traced open line with this surface.

## Data availability

The numerical data and analysis code that support the findings of this study are available from the corresponding author upon reasonable request.

## Acknowledgements

This work was supported by AFOSR under award number FA9550-25-1-0110.

## References

1. Scheeler, M. W., Kleckner, D., Proment, D., Kindlmann, G. L. & Irvine, W. T. M. Helicity conservation by flow across scales in reconnecting vortex links and knots. *Proc. Natl Acad. Sci. USA* **111**, 15350–15355 (2014).

2. Kleckner, D. & Irvine, W. T. M. Creation and dynamics of knotted vortices. *Nature Phys.* **9**, 253–258 (2013).

3. Moffatt, H. K. & Ricca, R. L. Helicity and the Călugăreanu invariant. *Proc. R. Soc. Lond. A* **439**, 411–429 (1992).

4. Călugăreanu, G. Sur les classes d'isotopie des noeuds tridimensionnels et leurs invariants. *Czech. Math. J.* **11**, 588–625 (1961).

5. Taylor, J. B. Relaxation of toroidal plasma and generation of reverse magnetic fields. *Phys. Rev. Lett.* **33**, 1139–1141 (1974).

6. Berger, M. A. Introduction to magnetic helicity. *Plasma Phys. Control. Fusion* **41**, B167–B175 (1999).

7. Berry, M. V. & Dennis, M. R. Knotted and linked phase singularities in monochromatic waves. *Proc. R. Soc. Lond. A* **457**, 2251–2263 (2001).

8. Dennis, M. R., King, R. P., Jack, B., O'Holleran, K. & Padgett, M. J. Isolated optical vortex knots. *Nature Phys.* **6**, 118–121 (2010).

9. Larocque, H. *et al.* Optical framed knots as information carriers. *Nature Commun.* **11**, 5119 (2020).

10. Bode, B. Stable knots and links in electromagnetic fields. *Commun. Math. Phys.* **387**, 1757–1770 (2021).

11. Sugic, D. *et al.* Particle-like topologies in light. *Nature Commun.* **12**, 6785 (2021).

12. Shen, Y. *et al.* Optical skyrmions and other topological quasiparticles of light. *Nature Rev. Phys.* **6**, 52–67 15–25).

13. Dennis, M. R. Topological singularities in wave fields. PhD thesis, University of Bristol (2001).

14. Adams, J. M., Agha, I. & Chong, A. Spatiotemporal optical vortex reconnections of multi-vortices. *Sci. Rep.* **14**, 4298 (2024).

15. Adams, J. M. & Chong, A. Spatiotemporal optical vortex reconnections of loop vortices. *Nanophotonics* (2025).

16. Adams, J. M. Spatiotemporal optical vortex phenomena. PhD thesis, University of Dayton (2024).

17. Zhou, Y. *et al.* Spatiotemporally localized optical links and knots. *arXiv*:2511.05908 (2025).

18. Ahlfors, L. V. *Complex Analysis: An Introduction to the Theory of Analytic Functions of One Complex Variable* 3rd edn (McGraw-Hill, 1979).

19. Guillemin, V. & Pollack, A. *Differential Topology* (Prentice-Hall, 1974).

20. Ricca, R. L. & Nipoti, B. Gauss' linking number revisited. *J. Knot Theory Ramifications* **20**, 1325–1343 (2011).

21. Milnor, J. *Singular Points of Complex Hypersurfaces* (Princeton Univ. Press, 1968).

22. Dennis, M. R. Local phase structure of wave dislocation lines: twist and twirl. *J. Opt. A: Pure Appl. Opt.* **6**, S202–S208 (2004).